# A Model of $f(R)$ Gravity as an Alternative for Dark Matter in Spiral Galaxies

Solmaz Asgari

Department of Science, Islamic Azad University, Abhar Branch

PO box 22, Abhar, Iran

E-mail: asgari_s@iau-abhar.ac.ir

Reza Saffari (Corresponding author)

Department of Science, the University of Guilan

PO box 41335-1914, Rasht, Iran

E-mail: rsk@guilan.ac.ir

**Abstract**

In this paper we study consistent solutions of spherically symmetric space in metric $f(R)$ gravity theory. Here we inversely obtain a generic action from metric solutions that describe flat rotation curves in spiral galaxies without dark matter. Then we show that obtained solutions are in conformity with Tully-Fisher relation and modified Newtonian dynamics, which are two strong constraints in justification of flat rotation curves in spiral galaxies.

**Keywords:** Rotation curve, Alternative gravity theories, Dark Matter

**1. Introduction**

During last century, General Relativity (GR) had unique successes in description of theory of gravity. The Standard Model of Cosmology (SMC) which established on GR drew a complete picture of cosmos. But recent cosmic observations revealed some realities that caused some scratches on SMC. Distance modulus-redshift data sets of SNeIa and cosmic microwave background (CMB) radiation indicate that universe is under a positive accelerating expansion phase (G. Hinshaw, 2007). Accepting of this reality depends on existence of a new type of energy and matter with extraordinary specifications which is called Dark Energy (DE) and Dark Matter (DM), respectively. On the other hand galactic scale and Solar system scale observations indicate some anomalous behaviors. Circular rotation curves of stars around the center of spiral galaxies, especially in large distances from the center, completely disagree with planetary dynamics that is based on Newtonian gravity. In Solar system, Pioneers 10 and 11 discovered an extra constant symmetric attractive acceleration toward Sun which has not any physical explanation. While there is no solution for these anomalies in GR, people propose DM to justify these anomalies (M. Persic, 1996). The idea of existence DE and DM which totally are about 96 percent of all components of energy and matter in the Universe is faced by many challenges such as assigning and explaining of essence and properties. In this open field, some works are based on making alternative gravity theories, to justify observational results without existence of DE and DM. The customary rule in this extending way is changing the geometric part of Einstein-Hilbert (EH) action from $R$ to a generic function such as $f(R)$, in which $R$ is Ricci scalar (S. Nojiri, 2003). In this paper following our previous works (R. Saffari, 2008), we suppose that gravitating behavior of cosmos is various in different scales and then we consider gravitating behavior of galactic scale to obtain an asymptotic behavior of a generic $f(R)$ model in this scale. Therefore we obtain consistent solutions of spherically symmetric space. The results of this approach in addition to justification of flat rotation curves in spiral galaxies, is in agreement with Tully-Fisher (TF) relation and is consistent with MOdified Newtonian Dynamics (MOND).

**2. Consistent solutions of spherically symmetric space**

Generic form of modified EH action is

$$S = \frac{1}{2\kappa} \int d^4 x \sqrt{-g} f(R) + S_m, \qquad (1)$$

where $\kappa$ is constant, $g$, is trace of diagonal metric and $S_m$ is matter action. Varying the action with respect to the metric results in the field equations as

$$F(R)R_{\mu\nu} - \nabla_\mu \nabla_\nu F(R) - \kappa T_{\mu\nu} = -g_{\mu\nu}(\frac{1}{2}f(R) - \nabla_\alpha \nabla^\alpha F(R)), \quad (2)$$

where $F(R) = df/dR$. Let us take a generic spherically symmetric metric as

$$ds^2 = -B(r)dt^2 + \frac{X(r)}{B(r)}dr^2 + r^2 d\Omega^2, \quad (3)$$

where $B(r)$ and $X(r)$ are radial components of metric. Whereas spherically symmetric space metric is diagonal, just there are two independent field equations in empty space as

$$\frac{X'}{X} = \frac{2rF''}{2F + rF'},$$
$$B'' + (\frac{F'}{F} - \frac{1}{2}\frac{X'}{X})B' - \frac{2}{r}(\frac{F'}{F} - \frac{1}{2}\frac{X'}{X})B - \frac{2}{r^2}B + \frac{2}{r^2}X = 0, \quad (4)$$

where the prime means $d/dr$. In the case of EH action ($F = 1$ and $X = 1$) Eqn. (4) reduces to Schwarzschild solution. In another word, $F = 1$ fix one of the functions amongst $F$, $X$ and $B$, then two differential equations are sufficient to solve this space. However for a genetic $f(R)$ to find a solution for the space three differential equations are needed. Thus to have a set of unique solutions for metric components we should define one of the three unknown functions before solving two differential equations. Then we propose a set of solutions as

$$F(r) = F_0 r^n,$$
$$X(r) = X_0 r^\alpha, \quad (5)$$
$$B(r) = B_0 r^\alpha,$$

where $F_0$, $X_0$, $B_0$, $n$ and $\alpha$ are constants which is bounded by two constraints as

$$\alpha = 2n(n-1)/(n+2),$$
$$X_0 = \left[1 + \frac{1}{2}n(2-\alpha) - \frac{1}{4}\alpha^2\right]B_0. \quad (6)$$

For metric space-time which introduced in Eq. (3), Ricci scalar defines as

$$R(r) = \frac{2}{r^2} + \frac{X'}{X^2}(\frac{1}{2}B' + \frac{2}{r}B) - \frac{1}{X}(B'' + \frac{4}{r}B' + \frac{2}{r^2}B), \quad (7)$$

and its value for solutions (5), obtains as

$$R(r) = \frac{2}{r^2} R_0(n,\alpha), \quad (8)$$

where

$$R_0(n,\alpha) = 1 - \frac{1 + \alpha/2 + \alpha^2/8}{1 + n(1 - \alpha/2) - \alpha^2/4}, \quad (9)$$

in which for $\alpha$ or $n \to 0$, $R_0$ and therefore $R \to 0$ which is Schwarzschild solution. Now by elimination of $r$ between $F(r)$ from (5) and $R(r)$ from (8) one can find an asymptotic form of $f(R)$ such as

$$f(R) = \frac{F_0 (2R_0)^{n/2}}{1 - n/2} R^{1-n/2} + \Lambda_0, \tag{10}$$

where $\Lambda_0$ is a constant of integration. If $n \to 0$ asymptotic solution for action goes to $F_0 R + \Lambda_0$ which forced our solution to define $F_0 = 1$ to agrees with GR. In this case we recovered GR with a cosmological constant. For $n = 2$ the action is undefined.

Consistency condition adds an extra equation to check the validity of solutions as

$$R'' f^{(2)}(R) + R'^2 f^{(3)}(R) = F''(r), \tag{11}$$

where $f^{(n)} \equiv d^n f / dR^n$. Any asymptotic solution of $f(R)$ which satisfy Eq. (11), makes a self-consistent system with (5) and (8). This consistency condition does not take consideration in (T. Multamaki, 2006) and (Y. Sobouti, 2007).

### 3. Application in rotation curve

In weak field approximation, geodesic equation for a test particle that rotates around the central mass obtains as $\ddot{r} + \Gamma^r_{tt} = 0$. Substituting the corresponding metric elements we get the following velocity for a particle rotating around the center of galaxy up to the fist order of $\alpha$ as

$$v^2 = \frac{1}{2} rc^2 \frac{B'_M(r) B_M(r)}{X(r)} = \frac{1}{2} c^2 \alpha B_0 r^\alpha, \tag{12}$$

Here for $\alpha \ll 1$ we expand $r^\alpha$ up to the first order of $\alpha$ as $r^\alpha \approx 1 + \alpha \ln r$ to get the asymptotic velocity of stars in large distances from the center of galaxy as

$$v_\infty^2 = \frac{1}{2} c^2 \alpha B_0, \tag{13}$$

In this solution asymptotic velocity depends on two parameters, $\alpha$ and $B_0$. $\alpha$ is related to $n$ by Eq. (6) which is the power of $R$ in the action (10), consequently it should be a small dimensionless number and independent of mass, because it comes from the geometric part of action. But in some works it is violated (Y. Sobouti, 2007). Then $B_0$ should be a dimensionless parameter. Here we describe that how $B_0$ relates to experimental Tully-Fisher relation as a constraint for our solutions. According to Tully-Fisher relation the forth power of asymptotic rotational velocity in large distance from the center of a spiral galaxy is proportional to mass of galaxy, $v_\infty^4 \propto M$. Here we notice that $\alpha$ is appeared in the geometric action and could not be dependent on mass, then we choose $B_0 = 2\sqrt{\mu M}$, in which $\mu$ is a proportional coefficient with mass inverse dimension and $M$ is mass of galaxy. Therefore asymptotic velocity obtains as

$$v_\infty^2 = c^2 \alpha \sqrt{\mu M_{galaxy}}, \tag{14}$$

in which $\alpha$ and $\mu$ will determine comparing with observational rotation curve data sets. But to have an estimation of these parameters for equation (14) we suppose mass of galaxy is about $10^{11} M_{Sun}$. For an asymptotic velocity which equals to $200\ kms^{-1}$ we obtain

$$\alpha^2 \mu = (v_\infty / c)^4 / M_{galaxy} \approx 10^{-54} kg^{-1}. \tag{15}$$

This value of $\alpha^2 \mu$ recovers Tully-Fisher's relation, but we should test it with another acceptable theory such as MOND results.

### 4. Equivalence with MOND

In the beginning of 80's decade, MOND theory introduced by Milgrom and obtained many successes in description of DM in spiral galaxies (M. Milgrom, 2002). Until 2004 MOND theory did not have a relativistic description, when Bekenstein introduced a rigorous Tensor-Vector-Scalar (TeVeS) theory for MOND paradigm (J. D. Bekenstein, 2005).

Gravitational acceleration in MOND theory obtains as $g = g_N$ for $g_N \gg a_0$ and $g = (a_0 g_N)^{1/2}$ for $a_0 \gg g_N$, where $g_N$ is Newtonian gravitational acceleration and $a_0 = 1.2 \times 10^{-10} ms^{-2}$, is MOND acceleration parameter.

In our solution, gravitational acceleration in weak field approximation up to the first order terms in $\alpha$ obtains as

$$g_M(r) = \left|\frac{d\phi_M(r)}{dr}\right| = \frac{v_\infty^2}{r} = c^2\alpha\frac{\sqrt{\mu M}}{r}, \qquad (16)$$

Therefore we can obtain the value of the composition of $\alpha^2\mu$ as

$$\alpha^2\mu = \frac{a_0 G}{c^4} = 0.98\times 10^{-54} \, kg^{-1}, \qquad (17)$$

which is in agreement of obtained value of the composition $\alpha^2\mu$ from asymptotic velocity for a typical galaxy.

## 5. Conclusion

The idea of application of $f(R)$ gravity theory at first introduced to justify late time acceleration of universe and early time inflation. But it is extended to cover all scales of theory of gravity. In this paper it is showed that modification of gravity by $f(R)$ for galactic scales is in complete agreement with two observational constraints such as rotation curve of spiral galaxies and Tully-Fisher relation. In this paper it is showed that a constant composition of model parameters as $\alpha^2\mu$ reaches to the same value due to comparing with observational data and MOND theory. According to obtained results from our previous works that calculates the value of parameter $\alpha \approx 10^{-6}$, then the value of another model parameter in this work, $\mu \approx 10^{-42} \ kg^{-1}$ which is very close to the inverse mass of a typical spiral galaxy.